\documentclass[superscriptaddress,aps,prl,twocolumn,showpacs,nofootinbib, longbibliography, notitlepage,floatfix]{revtex4-2}
\usepackage{etex}
\usepackage{amsmath,amssymb,amsthm}
\usepackage[colorlinks=true,citecolor=blue,urlcolor=blue]{hyperref}
\usepackage[pdftex]{graphicx}
\usepackage{times,txfonts}
\usepackage{braket}
\usepackage{color}
\usepackage{natbib}
\usepackage{amsmath,blkarray}
\usepackage{mathtools}
\usepackage{ulem}
\usepackage{latexsym}
\usepackage{tabularx, booktabs}
\usepackage{graphics,epstopdf}
\usepackage{graphicx}
\usepackage{float}
\usepackage{amsfonts}
\usepackage{tikz}
\usetikzlibrary{quantikz}
\usepackage{color,soul}

\newcommand{\be}{\begin{equation}}
\newcommand{\ee}{\end{equation}}
\newcommand{\ba}{\begin{eqnarray}}
\newcommand{\ea}{\end{eqnarray}}

\newcommand{\tr}{\operatorname{Tr}}

\usepackage{multirow}
\usepackage{appendix}
\usepackage{url}

\begin{document}
\title{Digitized-Counterdiabatic Quantum Optimization} 

\author{Narendra N. Hegade}
\email{narendrahegade5@gmail.com}
\affiliation{International Center of Quantum Artificial Intelligence for Science and Technology~(QuArtist) \\ and Physics Department, Shanghai University, 200444 Shanghai, China}

\author{Xi Chen}
\email{chenxi1979cn@gmail.com}
\affiliation{Department of Physical Chemistry, University of the Basque Country UPV/EHU, Apartado 644, 48080 Bilbao, Spain}

\author{Enrique Solano}
\email{enr.solano@gmail.com}
\affiliation{International Center of Quantum Artificial Intelligence for Science and Technology~(QuArtist) \\ and Physics Department, Shanghai University, 200444 Shanghai, China}
\affiliation{IKERBASQUE, Basque Foundation for Science, Plaza Euskadi 5, 48009 Bilbao, Spain}
\affiliation{Kipu Quantum, Kurwenalstrasse 1, 80804 Munich, Germany}

\date{\today}

\begin{abstract}
We propose digitized-counterdiabatic quantum optimization (DCQO) to achieve polynomial enhancement over adiabatic quantum optimization for the general Ising spin-glass model, which includes the whole class of combinatorial optimization problems. This is accomplished via the digitization of adiabatic quantum algorithms that are catalysed by the addition of non-stoquastic counterdiabatic terms. The latter are suitably chosen, not only for escaping classical simulability, but also for speeding up the performance. Finding the ground state of a general Ising spin-glass Hamiltonian is used to illustrate that the inclusion of k-local non-stoquastic counterdiabatic terms can always outperform the traditional adiabatic quantum optimization with stoquastic Hamiltonians. In particular, we show that a polynomial enhancement in the ground-state success probability can be achieved for a finite-time evolution, even with the simplest 2-local counterdiabatic terms. Furthermore, the considered digitization process, within the gate-based quantum computing paradigm, provides the flexibility to introduce arbitrary non-stoquastic interactions. Along these lines, using our proposed paradigm on current NISQ computers, quantum speed-up may be reached to find approximate solutions for NP-complete and NP-hard optimization problems. We expect DCQO to become a fast-lane paradigm towards quantum advantage in the NISQ era.
\end{abstract}

\maketitle

{\it Introduction.---} 
Many important optimization problems in science and industry can be formulated as solving combinatorial optimization problems \cite{fu1986application}. In general, these problems are known to be NP-hard, so that no classical or quantum computers are expected to solve them efficiently. However, there is a hope that quantum computers might give some polynomial speed-up, which helps reduce the resources and hence the cost of solving many practical problems of interest. Especially, adiabatic quantum optimization (AQO) algorithms are developed to tackle such problems \cite{farhi2001quantum, das2008colloquium, albash2018adiabatic}. Here, the solution to the optimization problem is encoded in the ground state of an Ising spin-glass Hamiltonian \cite{lucas2014ising}. The adiabatic theorem states that the system will remain in the instantaneous ground state if the evolution from the ground state of an initial Hamiltonian to the final Hamiltonian is sufficiently slow enough. The corresponding time-dependent Hamiltonian is given by
\begin{equation}\label{adb}
H_{ad}(\lambda)=\lambda(t)\left[\sum_{i<j} J_{i j} \sigma_{i}^{z} \sigma_{j}^{z}+\sum_{i} h_{i} \sigma_{i}^{z}\right]-(1-\lambda(t)) \left[ \sum_{i} \sigma_{i}^{x} \right ],
\end{equation}
where $\sigma^{z}$ and $\sigma^{x}$ are the Pauli operators. And, $\lambda(t) \in [0, 1]$ is a scheduling function represents the interpolation from the initial Hamiltonian to the final Hamiltonian. In Eq.~\eqref{adb}, the first term corresponds to the Ising spin-glass Hamiltonian with all-to-all interactions, finding its ground state in the worst case scenario is NP-hard \cite{barahona1982computational}. The second term with transverse field represents the initial Hamiltonian, corresponding to quantum fluctuation.

The Hamiltonian $H_{ad}(\lambda)$ has off-diagonal matrix elements that are real and non-positive in the computational basis, in other words, stoquastic and quantum Monte Carlo simulations can tackle such problems without facing any sign problem. It is believed that adiabatic quantum optimization or quantum annealing with stoquastic Hamiltonian might not give significant enhancement over classical algorithms, but some counterexamples have recently been found \cite{gilyen2021sub, hastings2021power, fujii2018quantum}. The adiabatic quantum computation with non-stoquastic Hamiltonians is known to be universal. However, the role of non-stoquastic catalysts to speed-up adiabatic quantum optimization problems is an unresolved problem. There are some problems where the non-stoquastic catalysts are advantageous \cite{farhi2009quantum, seki2012quantum, seki2015quantum, hormozi2017nonstoquastic, vinci2017non, nishimori2017exponential, outeiral2021investigating, takada2021phase}, and others are known to worsen the performance compared to their stoquastic counterparts \cite{crosson2020signing}. The main reason for this ambiguity is that the non-stoquastic terms are chosen randomly in all the previous works.

The counterdiabatic (CD) technique, borrowing from shortcuts to adiabaticity (STA)~\cite{sta4,guery2019shortcuts}, was introduced to speed up adiabatic evolutions by adding Hamiltonian terms to suppress the non-adiabatic transitions~\cite{demirplak2003adiabatic, berry2009transitionless, prlchen2011b,del2013shortcuts}. Recently, a number of developments have shown the advantage of CD techniques in digitized-adiabatic quantum computing~\cite{hegade2021shortcuts, DCQC2, DCQC3, DCQC4}, quantum annealing~\cite{CDAQC1, CDAQC2, hartmann2019rapid, prielinger2021two, CDAQC3}, and quantum approximate optimization algorithms (QAOA)~\cite{CDQAOA1, CDQAOA2, CDQAOA3}. In this manuscript, we propose digitized-counterdiabatic quantum optimization (DCQO) as a novel paradigm to solve the general class of combinatorial optimization problems with quantum speed-up. We show that the suitably designed CD terms appearing during the non-adiabatic evolution act as non-stoquastic catalysts. This provides us with a guaranteed enhancement over traditional adiabatic quantum optimization with stoquastic Hamiltonians, while solving the long-range Ising spin-glass problem. Moreover, we consider local approximate CD terms that can be obtained without knowing any prior detail of the Hamiltonian spectra~\cite{sels2017minimizing, claeys2019floquet, hatomura2021controlling}. We remark that we derive a general analytical expression for the CD coefficients, so that one does not have to calculate them explicitly for each case. Finally, we follow the gate-model approach for the digitized-adiabatic quantum evolution~\cite{barends2016digitized, hegade2021shortcuts}, allowing the digital implementation of arbitrary non-stoquastic CD terms in current noisy intermediate-scale quantum (NISQ) computers. Consequently, the proposed DCQO paradigm will be useful to solve the general class of combinatorial optimization problems with quantum speed-up in the NISQ era.

{\it Counterdiabatic driving as a non-stoquastic catalyst.---} The concept of STA was originally proposed last decade \cite{sta4}, and was found to have wide applications in many fields, ranging from quantum physics, classical physics to stochastic physics \cite{guery2019shortcuts}. Among all techniques of STA, CD driving, also called transitionless driving, provided the possibility of tailoring the Hamiltonian of quantum many-body systems \cite{sels2017minimizing, claeys2019floquet} to speed up the adiabatic process.  Here, the source adiabatic Hamiltonian is added by a CD term that takes the form
\begin{equation}\label{totalHam}
    H(\lambda) = (1- \lambda(t))H_{i} + \lambda(t)H_{f}  + f(\lambda) H_{cd} .
\end{equation}
Here, $H_{i}$ and $H_{f}$ are the initial and the problem Hamiltonians connected by $\lambda(t)$, satisfying $\lambda(0)=0$ and $\lambda(T)=1$, and $H_{cd}$ is the CD term with scheduling $f(\lambda)$, which vanishes at the beginning and end of the protocol. In principle, introducing the exact CD term help to follow the instantaneous ground state of the original Hamiltonian $H_{ad}(\lambda)$ at all times during the evolution. This motivates us to define the CD term as $H_{cd}(\lambda) = \dot{\lambda}A_{\lambda}$. We can notice that, when the evolution is very fast, $H_{cd}$ increases dramatically, and in the adiabatic limit, i.e., $|\dot{\lambda}| \rightarrow 0$, the CD term vanishes. Here, $A_{\lambda} = i \sum_m (1-\ket{m(\lambda)} \bra{m(\lambda)})\ket{\partial m(\lambda)}\bra{m(\lambda)}$ is known as adiabatic gauge potential corresponding to the non-adiabatic transitions between the eigen states $\ket{m(\lambda)}$ \cite{kolodrubetz2017geometry}, which satisfies the condition $\left[i \partial_{\lambda} H_{ad}-\left[{A}_{\lambda}, H_{ad}\right], H_{ad}\right]=0$. Obtaining the exact $A_{\lambda}$ for a many-body system is challenging. Its implementation is not optimal because, in many cases, $A_{\lambda}$ can contain exponentially many terms with non-local many-body interactions. An alternative approach is to consider the approximate form of the adiabatic gauge potential \cite{sels2017minimizing, hatomura2021controlling} that can be obtained from a nested commutator (NC) \cite{claeys2019floquet}  
\begin{equation}
    A_{\lambda}^{(l)} = i \sum_{k = 1}^l \alpha_k(t) \underbrace{[H_{ad},[H_{ad},......[H_{ad},}_{2k-1}\partial_{\lambda} H_{ad}]]].
    \label{gauge}
\end{equation}
Here, $\alpha_k(t)$ is the CD coefficient, obtained by minimizing the operator distance between the exact gauge potential and the approximate gauge potential. This is similar to minimizing the action $S =\tr[G_{\lambda}^2]$, where the Hilbert-Schmidt operator $G_{\lambda} = \partial_{\lambda} H_{ad} + i [A_\lambda^{(l)}, H_{ad}]$. For the real-valued Hamiltonian in Eq.~\eqref{adb}, the adiabatic gauge potential is always imaginary, so the terms appearing in the NC expansion always contain an odd number of Pauli-$y$ terms. So, by restricting to lower-order terms obtained from Eq.~\eqref{gauge} and postselecting only one-spin and two-spin interaction terms, one can construct the general 2-local CD term as
\begin{equation}\label{2LCD}
H_{cd}(\lambda)=\sum_{i} \alpha_{i}(\lambda) \sigma_{i}^{y}+\sum_{i \neq j} \beta_{i j}(\lambda) \sigma_{i}^{z} \sigma_{j}^{y}+\gamma_{i j}(\lambda) \sigma_{i}^{x} \sigma_{j}^{y} \, ,
\end{equation}
where the CD coefficients $\alpha_{i}$, $\beta_{i j}$, and $\gamma_{i j}$ is obtained by variational minimization \cite{sels2017minimizing, gentini2021variational}. We can notice that, by construction, the approximate CD terms have imaginary numbers as the off-diagonal matrix elements, which makes it non-stoquastic. So, the NC ansatz in Eq.~\eqref{gauge} serves as a prescription for choosing the non-stoquastic catalyst. However, not all the terms in the NC ansatz give significant enhancement. Therefore, preselecting the correct operators can help reduce the cost of implementation. Besides, by introducing more control parameters, the use of machine learning techniques and quantum variational algorithms for obtaining optimal values might further enhance the performance \cite{CDQAOA1, yao2020noise, gentini2021variational, hegde2021genetic, kadowaki2021greedy}.

We start with a simple local adiabatic gauge potential, which takes the form $\Tilde{A}_{\lambda}=\sum_{i}^N \beta_{i}(t) \sigma_{i}^{y}$, where the general expression for the CD coefficient $\beta_{i}(t)$ is calculated as $\beta_i(t) = h_i/{2 \left[(\lambda -1)^2+\lambda^2 \left({h_i}^2+ \sum_{i}J_{ij}^2\right)\right]}$. Even though the off-diagonal matrix elements for $\Tilde{A}_{\lambda}$ contains complex values, with a simple change of basis, one can make it stoquastic. Also, from the first order expansion of the NC ansatz we obtain the 2-local CD term as 
\begin{align}
\label{ACD}
    H_{cd}^{(1)}(\lambda) =- 2 \dot{\lambda} \alpha_1(t) \left[ \sum_i h_i {\sigma_i^y} + \sum_{i<j} J_{ij} (\sigma_i^y\sigma_j^z + \sigma_i^z \sigma_j^y)\right].
\end{align}
Here, the CD coefficient $\alpha_1(t) = -\frac{1}{4} \left[ \sum_i h_i^2 +2 \sum_{i<j} J_{ij}^2\right] /R(t)$, where $R(t)$ is given by
\begin{align}
\label{Eq6}
R(t) &= (1-2\lambda(t))\left[\sum_i h_i^2 +8 \sum_{i<j} J_{ij}^2\right] + \lambda(t)^2 \left[\sum_i h_i^2 + \sum_i h_i^4 \right. \nonumber \\ & \left .+8 \sum_{i<j} J_{ij}^2  + 2 \sum_{i<j} J_{ij}^4 + 6 \sum_{i\neq j} h_i^2 J_{ij}^2 +  6 \sum_{i< j}\sum_{k< l} J_{ij}^2 J_{kl}^2\right]. 
\end{align}
In the last term of Eq.~(\ref{Eq6}), we have the following additional constraints: $i=k$ or $j=l$, and equivalently $i=l$ or $j=k$. In Eq.~(\ref{ACD}), we only have a single variational parameter $ \alpha_1(t) $. For better performance, one can also consider the CD term in Eq.~\eqref{2LCD}, which requires optimizing $N^2$ parameters considering $\beta_{ij} = \beta_{ji}$, and $\gamma_{i j} = \gamma_{ji}$. For convenience, we use the shorthand notation, $Y =\dot{\lambda} \sum_{i}^N \beta_{i}(t) \sigma_{i}^{y}$, and $Y|ZY = H_{cd}^{(1)}(\lambda)$.

{\it Digitized counterdiabatic driving.---}
For adiabatic quantum optimization problems, the CD terms are non-stoquastic, making it challenging to realize such Hamiltonian on existing quantum annealers \cite{ozfidan2020demonstration, vinci2017non}. Also, for obtaining higher success probability in many-body systems, it is essential to consider higher-order k-local CD terms. This is a challenging task for analog quantum computers and quantum annealers. Along with that, the lack of flexibility in realizing arbitrary interactions is one of the main drawbacks of analog quantum computers and quantum annealers. To overcome all these problems, we used digitized-adiabatic quantum computing techniques~\cite{barends2016digitized, hegade2021shortcuts}, which provides the flexibility to introduce arbitrary multi-qubit and non-stoquastic interactions. The model is also consistent with error correction \cite{fowler2012surface}, and error mitigation techniques are being developed for NISQ computers~\cite{endo2018practical}.

\begin{figure}
    \centering
    \includegraphics[width=\linewidth]{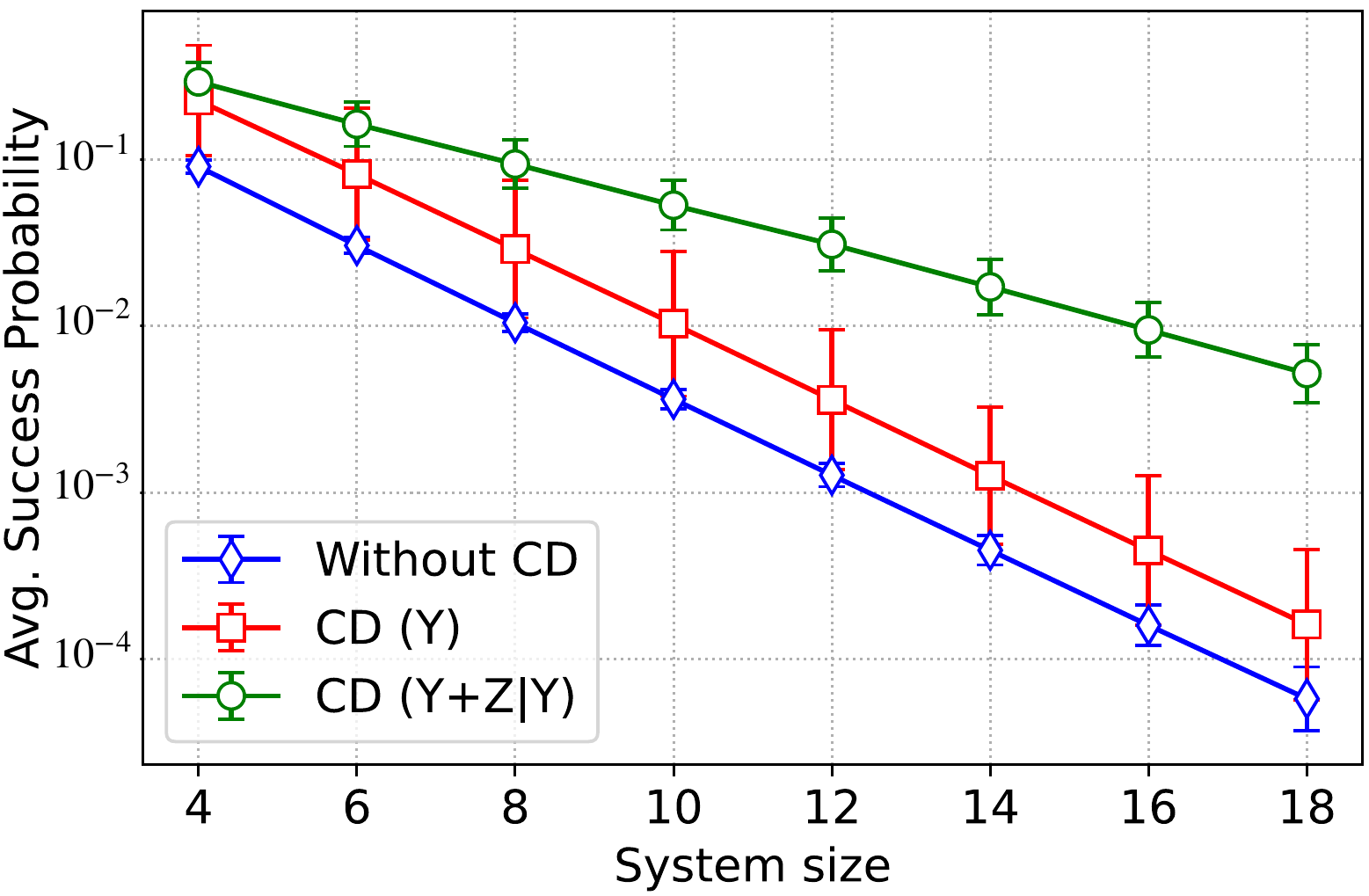}
    \caption{The Average success probability of obtaining the ground state as a function of number of spins in the Ising spin-glass Hamiltonian is depicted. Here, we fix the total evolution time $T=1$ and the number of trotter steps to 20, \textit{i.e.} $\delta t = 0.05$. The interaction strengths and the local fields are chosen randomly from a Gaussian distribution for 1000 random instances. The blue curve corresponds to the evolution without the CD term, the red curve is with the local single qubit CD term, and the green curve is for the 2-local CD term in Eq.~\eqref{ACD}. In the latter case, a polynomial enhancement in the success probability can be observed.}
    \label{fig1}
\end{figure}

\begin{figure*}
    \centering
    \includegraphics[width=\linewidth]{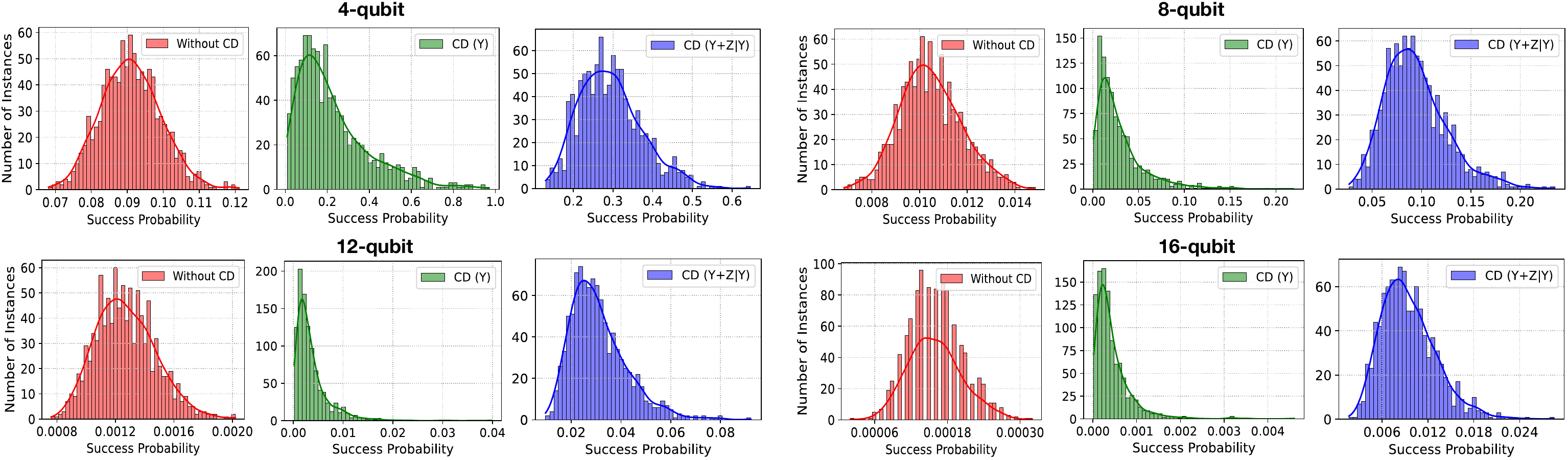}
    \caption{The distribution of ground-state success probability for 1000 randomly chosen instances are illustrated. The red plot corresponds to the traditional stoquastic adiabatic Hamiltonian in Eq.~\eqref{adb}, the green plot is for the local single-spin CD term, and the blue plot is for the non-stoquastic 2-local CD term in Eq.~\eqref{ACD}.}
    \label{fig2}
\end{figure*}

For the time-dependent Hamiltonian in Eq.~\eqref{totalHam}, the evolved state is given by $\ket{\psi(T)} = \mathcal{T} e^{-i \int_{0}^{T} H(\lambda) dt}\ket{\psi(0)}$, where $\ket{\psi(0)} = \frac{1}{\sqrt 2}(\ket{0}+\ket{1})^{\otimes N}$, and $\mathcal{T}$ is the time ordering operator. The total Hamiltonian can be decomposed into sum of local terms, \textit{i.e.}, $H(\lambda) = \sum_j \gamma_j(t) H_j$. We discretize the total time $T$ into many small intervals of size $\delta t$. Using the first order Trotter-Suzuki formula, we obtained the approximate time evolution operator given by
\begin{equation}
    U_{\mathrm{dig}}(0, T) = \prod_{k=1}^{M} \prod_{j} \exp \left\{-i  \delta t  \gamma_j(k \delta t) H_{j}\right\},
    \label{evolution}
\end{equation} 
where $M$ corresponds to the number of trotter steps. For better approximation one can also consider the recently proposed commutator product formulas \cite{chen2021efficient}. For the evolution of a time-dependent Hamiltonian, the trotter step size $\delta t$ should be less than the fluctuation time scale of the Hamiltonian \cite{poulin2011quantum}, \textit{i.e.}, $\delta t \ll\|\partial H / \partial t\|^{-1}$. However, recently, it has been shown that the digitized-adiabatic evolution is robust again discretization error \cite{yi2021robustness}. This loosens the restrictions on $\delta t$. The cost associated with the digitized-counterdiabatic evolution is given by $\text{Cost} = T \max _{\lambda}\|H(\lambda(t))\|$, which corresponds to the total gate count, while the number of trotter steps decide the circuit depth. For the fully-connected Ising spin-glass problem, total $N(N-1)/2$ entangling operations are needed for each trotter step, and the inclusion of the 2-local CD terms increases it by a constant factor.
Recently, it has been shown that implementing such fully connected Ising spin-glass problems on current quantum annealers, and also on parity quantum computers, results in huge time overhead because of the embedding schemes~\cite{embeddingissue}. In this regard, it is argued that gate-model quantum computing on a 2-D grid has an advantage compared to other architectures. In this sense, trapped-ion systems with all-to-all connectivity would be an ideal choice but they are not strictly necessary. 

In our simulation, we fix the total time $T=1$, and the step size $\delta t = 0.05$. The scheduling function is chosen as $\lambda(t)=\sin ^{2}\left[\frac{\pi}{2} \sin ^{2}\left(\frac{\pi t}{2 T}\right)\right]$, so that CD terms vanish at the beginning and the end of the evolution. Each matrix exponential term in Eq.~\eqref{evolution} is implemented using standard two-qubit CNOT gates and single-qubit rotations. For all the cases, the number of shots is chosen between 10000-100000 depending on the system size. By measuring the qubits in the computational basis, we obtain the success probability given by $P_s=\left| \braket{\psi_g| \psi_f(\lambda = 1)} \right |^2 $. Here, $\ket{\psi_g}$ is the actual ground state, and $\ket{\psi_f(\lambda = 1)}$ is the time evolved state at $t=T$. For the Hamiltonian in Eq.~\eqref{adb}, the coupling terms $J_{ij}$ and the local fields $h_i$ are chosen randomly from a continuous Gaussian distribution with unit variance and zero mean. 
For estimating the fraction of instances where the inclusion of the CD term gives an enhancement, we define a metric called enhancement ratio, given by $R_{enh} = L^{cd}/L$. Here, $L^{cd}$ is the number of instances with enhanced performance by including the CD term, and $L$ is the total number of instances, where we set $L=1000$. In order to quantifying the improvement in the success probability, we define a quantity called success probability enhancement, given by $P_{enh} = P_s^{cd}/P_s^{ad}$. Here, $P_s^{cd}$ and $P_s^{ad}$ are the success probability with and without the CD terms, respectively.

\begin{figure}
    \centering
    \includegraphics[width=\linewidth]{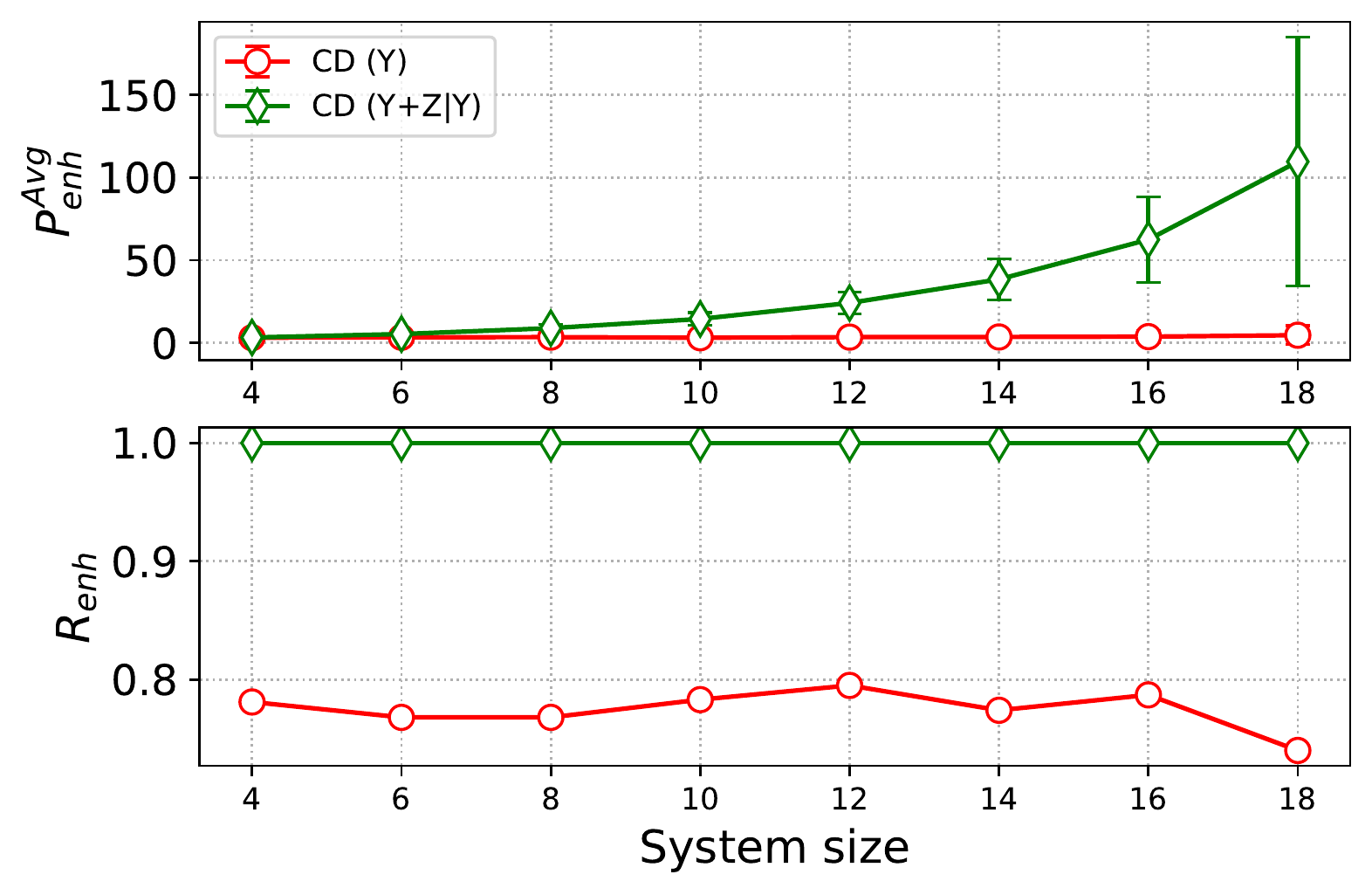}
    \caption{The average success probability enhancement ($P^{Avg}_{enh}$) and probability enhancement ratio ($R_{enh}$) as a function of system size for the Ising spin-glass problem is depicted. For the 2-local CD term from the first order NC ansatz, $P^{Avg}_{enh}$ increases with the system size. Whereas for the local CD term ($Y$), a constant enhancement by a factor of 3 is observed. In the bottom panel, we see that, the 2-local CD term always give enhancement for all the 1000 random instances, whereas the local CD has an average enhancement ratio ~0.756.}
    \label{fig3}
\end{figure}

\begin{figure}
    \centering
    \includegraphics[width=\linewidth]{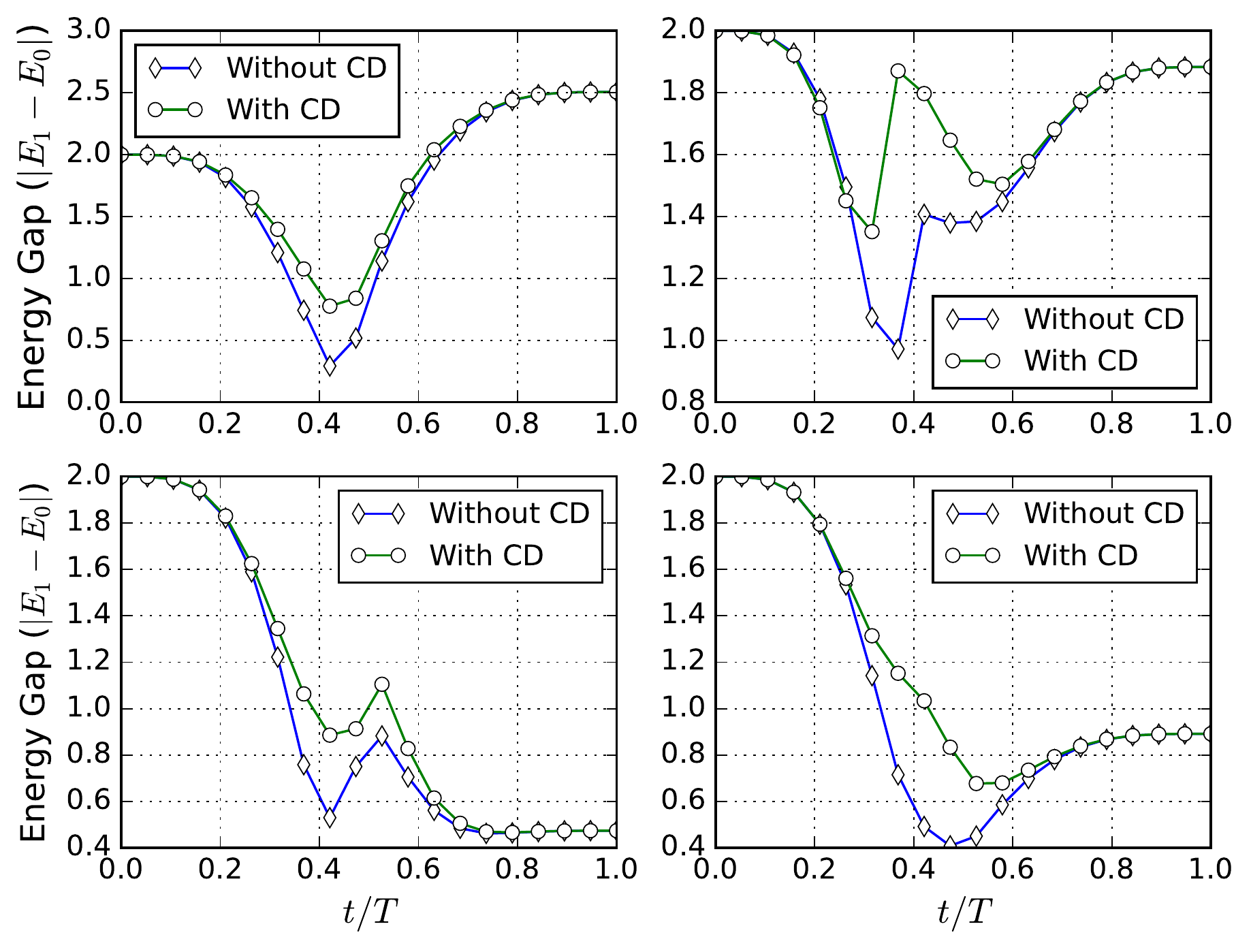}
    \caption{The energy gap between the ground state and the first excited state is plotted as a function of time for a system size $N=10$. The blue curve corresponds to the stoquastic Hamiltonian in Eq.~\eqref{adb}, while the green curve corresponds to the non-stoquastic Hamiltonian by including the 2-local CD term in Eq.~\ref{ACD}.} 
    \label{fig4}
\end{figure}

In Fig.~\ref{fig1}, the average ground-state success probabilities for the naive stoquastic Hamiltonian with only transverse field in Eq.~\eqref{adb}, including the local CD term $Y$ and 2-local CD term obtained from NC ansatz $Y|ZY$, are depicted for system size up to 18 qubits. We see that the success probability decreases rapidly with increasing system size for the naive non-adiabatic approach, the inclusion of the 2-local CD term $H_{CD}^{(1)}$ gives a polynomial enhancement, and the local single-spin CD gives a constant enhancement. In Fig.~\ref{fig2}, the histogram shows the ground-state success probability distribution for evolution with and without including the CD term for 1000 random instances. The top panel of Fig.~\ref{fig3} shows the average success probability enhancement ($P_{enh}^{Avg}$) by including the CD terms $Y$ and $Y|ZY$. As the system size grows, $P_{enh}^{Avg}$ increases polynomially for the CD term obtained from the first order NC ansatz. With the single spin CD term $Y$, we obtained an enhancement by a factor of $\approx3$, irrespective of the system size. The bottom panel of Fig.~\ref{fig3} depicts the fraction of instances where the inclusion of the CD term gives enhancement. For the local CD $Y$, we obtained $R_{enh}\approx 75.6\%$, while the CD term $Y|ZY$ gives the enhancement ratio $R_{enh}\approx 100\%$, indicating that the 2-local CD term gives a guaranteed enhancement for all the random instances. In general, the inclusion of CD terms does not help reduce the minimum gap, and its enhancement is explained by the fact that the additional terms help suppress the matrix elements responsible for the excitations between the eigenstates. However, to analyse the effect of non-stoquastic CD terms on the minimum gap $\Delta_{min}$, we study the instantaneous energy spectrum as a function of time. Surprisingly, we noticed that the inclusion of approximate CD terms obtained from the NC ansatz increases the minimum energy gap between the ground state and the first excited state during the evolution. This increased gap helps to reduce the excitations, resulting in increased success probability. In Fig.~\ref{fig4}, the energy gap between the ground state and the first excited state ($|E_1 - E_0|$) as a function time is plotted for four randomly chosen instances with system size $N=10$.

{\it Discussion and conclusion.---}
We have answered a long-debated problem in adiabatic quantum optimization, i.e., the speed-up role of non-stoquastic catalysts. We showed that cleverly chosen non-stoquastic counterdiabatic Hamiltonians achieve enhanced performance compared to traditional stoquastic adiabatic methods. We considered the general Ising-spin glass Hamiltonian with all-to-all connectivity to show that a polynomial enhancement in the ground-state success probability can be obtained, even with 2-local non-stoquastic CD terms stemming from the NC ansatz. As an outlook, considering higher-order k-local CD terms may further enhance the already observed quantum speed up of DCQO paradigm. In this work, we also provided a general analytical expression for scheduling these CD terms, whose calculation does not require any prior knowledge of the Hamiltonian spectra or the structure of the eigenstates. In conclusion, we proved that the proposed digitized counterdiabatic quantum optimization (DCQO) paradigm involving suitable non-stoquastic CD terms is superior to the traditional adiabatic quantum optimization (AQO) using stoquastic Hamiltonians. In this sense, DCQO might help to achieve quantum advantage for obtaining approximate solutions to combinatorial optimization problems on noisy intermediate-scale quantum (NISQ) computers from hundreds to few thousand qubits.

\begin{acknowledgments}
{\it Acknowledgments.} This work is supported by NSFC (12075145), STCSM (2019SHZDZX01-ZX04), Program for Eastern Scholar, EU FET Open Grants Quromorphic (828826) and EPIQUS (899368). X. C. acknowledges the Ram\'on y Cajal program (RYC-2017-22482).
\end{acknowledgments}

\bibliography{reference.bib}
\end{document}